\documentclass[a4paper,11pt]{article}
\pdfoutput=1 

\usepackage{jinstpub} 

\usepackage{lineno}

\usepackage{hyperref}

\title{\boldmath Compact multi-channel analyzer for SiPM detectors with real time on-board signal analysis}

\author[a,1]{P. Kučera,\note{Corresponding author.}}
\author[a]{O. Urban,}
\author[a]{O. Růžička,}
\author[a]{M. Vítek,}
\author[a]{P. M. Stašek,}
\author[a,d]{M. Holík,}
\author[a]{R. Klesa,}
\author[b,c]{F. Ahmadov,}
\author[b,c]{A. Sadigov,}
\author[b]{A. Mammadli,}
\author[b]{Ch. Abbasova,}
\author[d]{O. Pavlas}
\author[d]{and T. Slavíček}

\affiliation[a]{University of West Bohemia,\\Faculty of Electrical Engineering,
                \\ Univerzitní 8, 301 00 Plzeň, Czech Republic}
\affiliation[b]{Azerbaijan National Academy of Sciences,
                \\Institute of Radiation Problems,
                \\ANAS, B.Vahabzade str. 9, Baku, Azerbaijan}
\affiliation[c]{National Nuclear Research Center,
                \\Inshaatchilar str. 4, Baku, Azerbaijan}
\affiliation[d]{Czech Technical University,
                \\Institute of Experimental and Applied Physics,
                \\Husova 240/5 110 00 Prague 1, Czech Republic}

\emailAdd{pkucera@fel.zcu.cz}

\abstract{Multi-channel analysers (MCAs) play a crucial role in spectral measurements, especially in the context of Silicon Photomultipliers (SiPMs) used for gamma spectroscopy. Most commercial MCAs, while highly accurate, remain cost-prohibitive for broader applications. This paper presents the design and implementation of a cost-effective MCA utilizing off-the-shelf components while achieving spectroscopy of gamma particles with reasonable resolution. The MCA board is built around the STM32G4 family microcontroller (MCU), which provides embedded analog components, timers, and high-resolution ADCs. This system is designed to reduce external component requirements, thereby decreasing costs and increasing system reliability. Experimental results demonstrate that the MCA can perform accurate real-time gamma radiation measurements with SiPM detectors. The device offers flexible connectivity options (USB, Ethernet, WiFi). The low-cost, accessible nature of the MCA opens up opportunities for educational and research applications in radiation detection.}


\begin{document}
\maketitle
\flushbottom

\section{Introduction}
\label{sec:intro}
The use of SiPM detectors has become more widespread in radiation detection due to their high sensitivity, compact size~\cite{tancioni2021gammadoseratemonitoring,Buonanno2023,david_p__mcelroy__2007,ahmadov2022investigation}. They are smaller and operate at a lower voltage compared to photomultiplier tubes (PMTs).~\cite{Holik2020_read-out}.  However, MCAs that interface with these detectors remain largely inaccessible to researchers and educators due to their high costs and complexity because of using FPGA or ASIC~\cite{paolo_trigilio__2018,Holik2020_read-out}. This paper proposes a compact, cost-effective MCA system that retains the accuracy required for SiPM read-out. By using the STM32G4 microcontroller family, the design is optimized for cost-efficiency while maintaining performance through on-board signal processing capabilities. The primary goal of this work was to develop a read-out interface device for SiPMs and SiPM-based detectors with a focus on MAPD (Micropixel Avalanche
PhotoDiode) detector family~\cite{SiPM:SADYGOV200670}, which is available to a wider audience, including academic institutions and small-scale research labs.

\section{MCA Hardware Implementation}
The block diagram of the developed MCA is shown in figure ~\ref{fig:Block-diagram} below. It consists of three signal channels, MCU, bias source and various communications interfaces. Each part of the MCA is described in the subsection below and the real photo of the device is at the end of this section on the left part in figure ~\ref{fig:boards-photo}.

\begin{figure}[htbp]
\centering 
\includegraphics[width=0.95\textwidth,trim=0 10 0 0,clip]{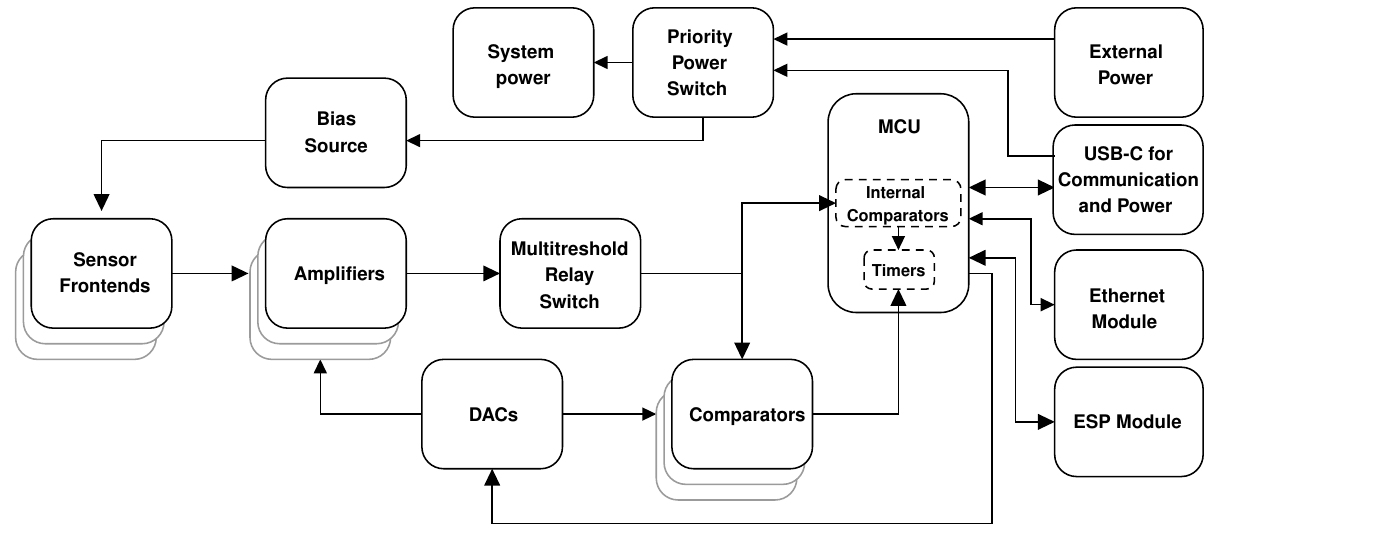}
\qquad
\caption{\label{fig:Block-diagram} Block diagram of the developed MCA with main parts of the signal path}
\end{figure}

\subsection{Microcontroller}
The STM32G474 was selected as the main microcontroller for its affordability and extensive analog functionality. It features an Arm® Cortex®-M4 32-bit MCU, digital interfaces, and low-power modes. Key features include high-resolution timers, ADCs, DAC channels, analog comparators, and operational amplifiers, making it ideal for developing a nearly single-chip MCA for SiPM detectors. Additional signal processing was added to this board for performance enhancement and comparison of internal analog blocks with external components. Future versions may simplify the MCA if internal analog performance proves sufficient.

\subsection{Sensor Frontend}
The MCA board supports up to three detector modules in single-threshold mode or one module in multi-threshold mode. Detector modules are connected via RJ45 connectors and standard twisted-pair cables (UTP/FTP), simplifying installation. One cable pair connects to the SiPM diode, and another to power (3.3~V) for temperature measurement and detector identification via EEPROM with a unique identifier using I2C. A third pair handles I2C communication, and the fourth is used for addressing. Each detector has a bias voltage with heavy RC filtering for low-ripple, precision measurement.



\subsection{Signal Amplifier and Fast External Comparator}
The SiPM signal, characterized by a small amplitude, is amplified using the LTC6252 operational amplifier, selected for its high bandwidth (720~MHz), fast slew rate (280~V/µs), low noise, and rail-to-rail output. The amplifier is configured as an inverting amplifier with a gain of 48, and the circuit's input resistor is matched to the SiPM’s impedance. Offset voltage regulation is achieved using a 12-bit external DAC, an integrated microcontroller DAC, or a trimmer resistor.

For signal evaluation, both internal and external comparators are employed to compare their performance in precise measurements. If the internal comparators demonstrate sufficient performance, the external comparators may be eliminated in future designs to reduce cost. The output signals of both comparators are routed with length-matched traces to the SMA connectors

\subsection{Powering Topology}
The board supports two power options: USB Type-C (5~V) or an external DC source (10-24V) via a DC-DC buck converter. A power management IC switches between power sources, prioritizing the external DC. The 5V rail is regulated down to 3.3~V for the microcontroller and circuitry, while also powering the bias supply.

\subsection{Bias Supply}
A switch-mode boost converter (LM51561-Q1) generates the bias voltage (40-80~V) with a high switching frequency (250~kHz) and heavy RC filtering for low ripple. The output voltage is monitored by the microcontroller via ADC and controlled  DAC, with a current injection technique for voltage adjustment, see current injection technique in~\cite{TI:current_injection}.

\subsection{Communication}
The MCA offers three communication options: USB Type-C (UART to USB via FTDI), Ethernet (10BaseT/100BaseTX for TCP/UDP), and Wi-Fi via an onboard ESP32 module. The Wi-Fi feature allows users to access data remotely via a smartphone. This feature can be used for quick and easy access to the measured data in the field during measurements.

\subsection{Longevity of Off-the-Shelf Components}
The MCA design leverages off-the-shelf components for cost-effectiveness and ease of procurement, with their typical lifespans being sufficient for our purposes. The STM32G474 microcontroller from STMicroelectronics, backed by a 10-year availability guarantee, ensures reliable long-term support. Components like operational amplifiers and comparators can be seamlessly replaced with compatible alternatives sharing the same package and pinout. Power sources, with a standard lifespan of 10 years, align well with the system’s operational needs and can be updated as required. Although the ESP32 module used for Wi-Fi communication is now obsolete, it can be replaced in future iterations without impacting overall functionality or performance.

\begin{figure}[htbp]
\centering 
\includegraphics[width=0.7\textwidth,trim=10 0 10 0,clip]{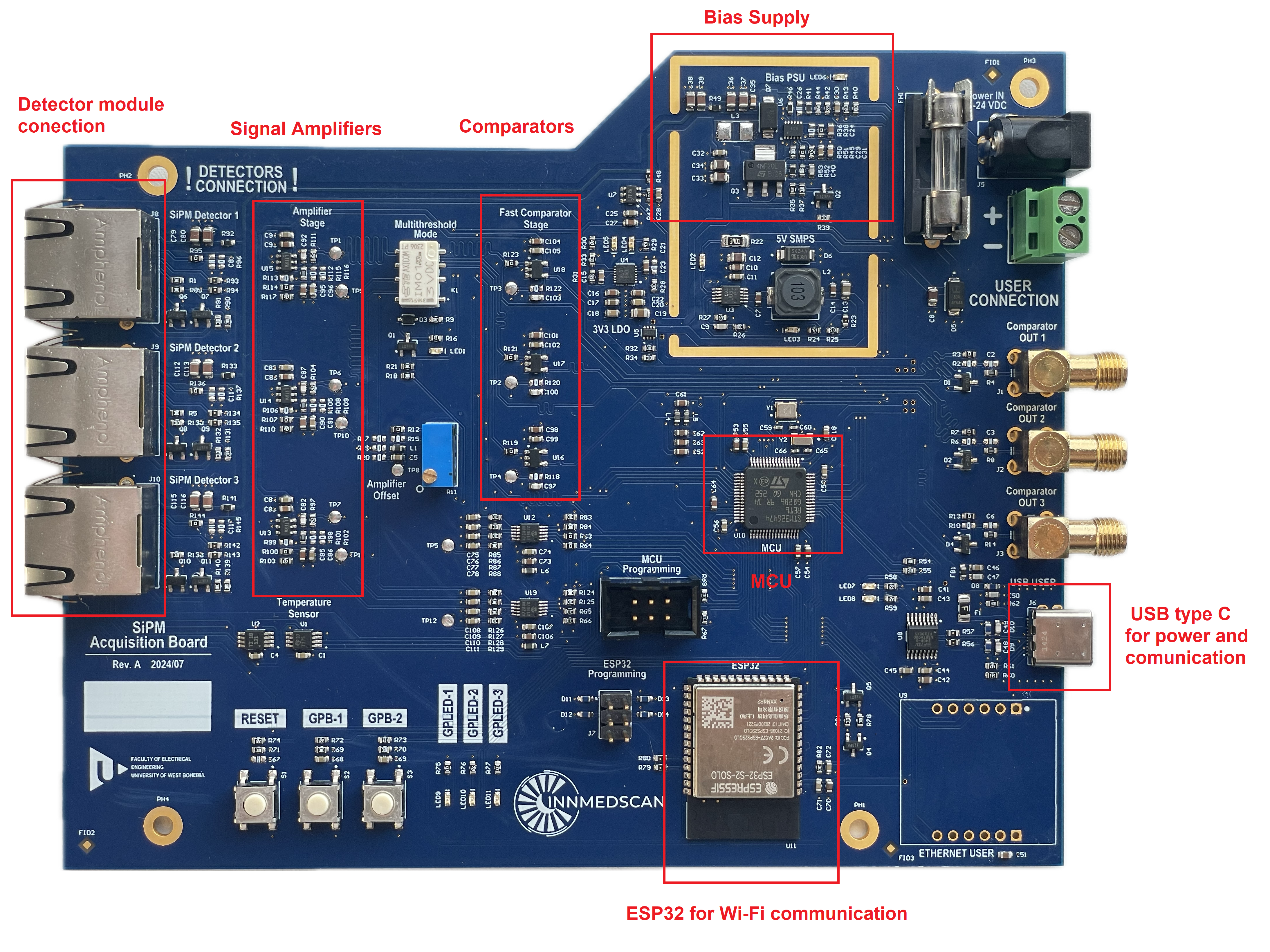}
\qquad
\includegraphics[width=0.2\textwidth,trim=10 -3000 10 0 0,clip]{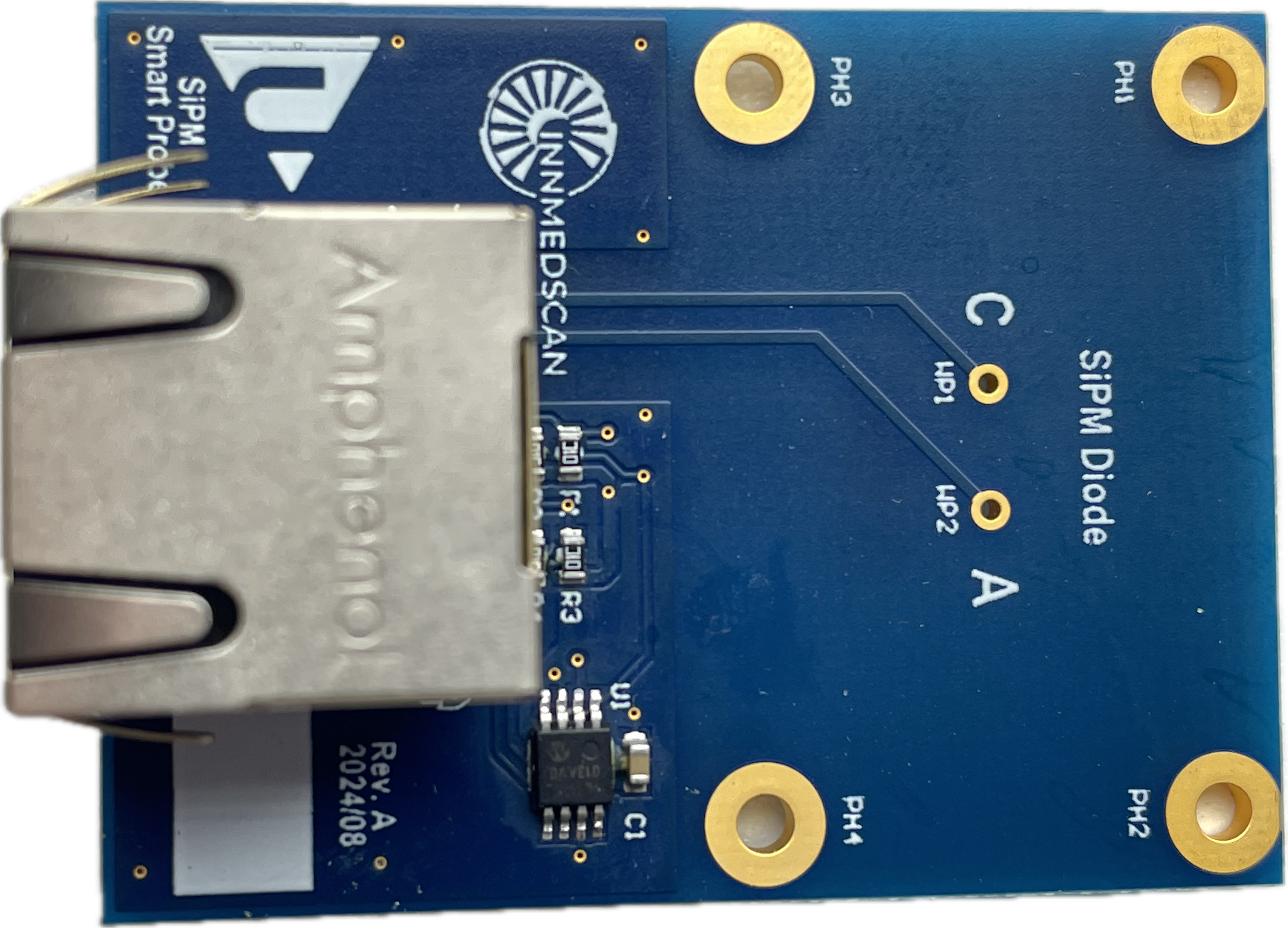}
\caption{\label{fig:boards-photo} Top view on the MCA device - the main functional parts on the board are marked (left), Detector module for connecting SiPM with temperature sensor and EEPROM for a unique identifier (right)}
\end{figure}

\section{Detector Module Hardware Implementation}
The detector module is designed for simplicity and low-cost manufacturing, as it will be mounted in high-radiation areas. Measuring only 35 x 46.5 mm, it directly couples to the scintillator and connects to the MCA via a single shielded UTP or equivalent cable, making installation economical and convenient. The detector module is shown on the right side in figure ~\ref{fig:boards-photo}.

The SiPM diode is soldered directly to the PCB, with a temperature sensor located underneath. The sensor and an EEPROM for storing detector ID and calibration values communicate via the I2C bus. Their addresses are set by hardware pins connected to the MCA port. Since radiation can degrade active components, the MCA includes a power switch to turn off the sensor and memory when not in use, limiting degradation.


\section{Acquisition software}
The acquisition software configures and controls data collection from the MCA, offering various user options for setup. Users can select the processing method, described in section ~\ref{sec:ToTMeasurement}, and control the SiPM bias voltage with separate displays for set and actual values.

\begin{figure}[htbp]
\centering 
\includegraphics[width=0.8\textwidth,trim=0 0 0 0,clip]{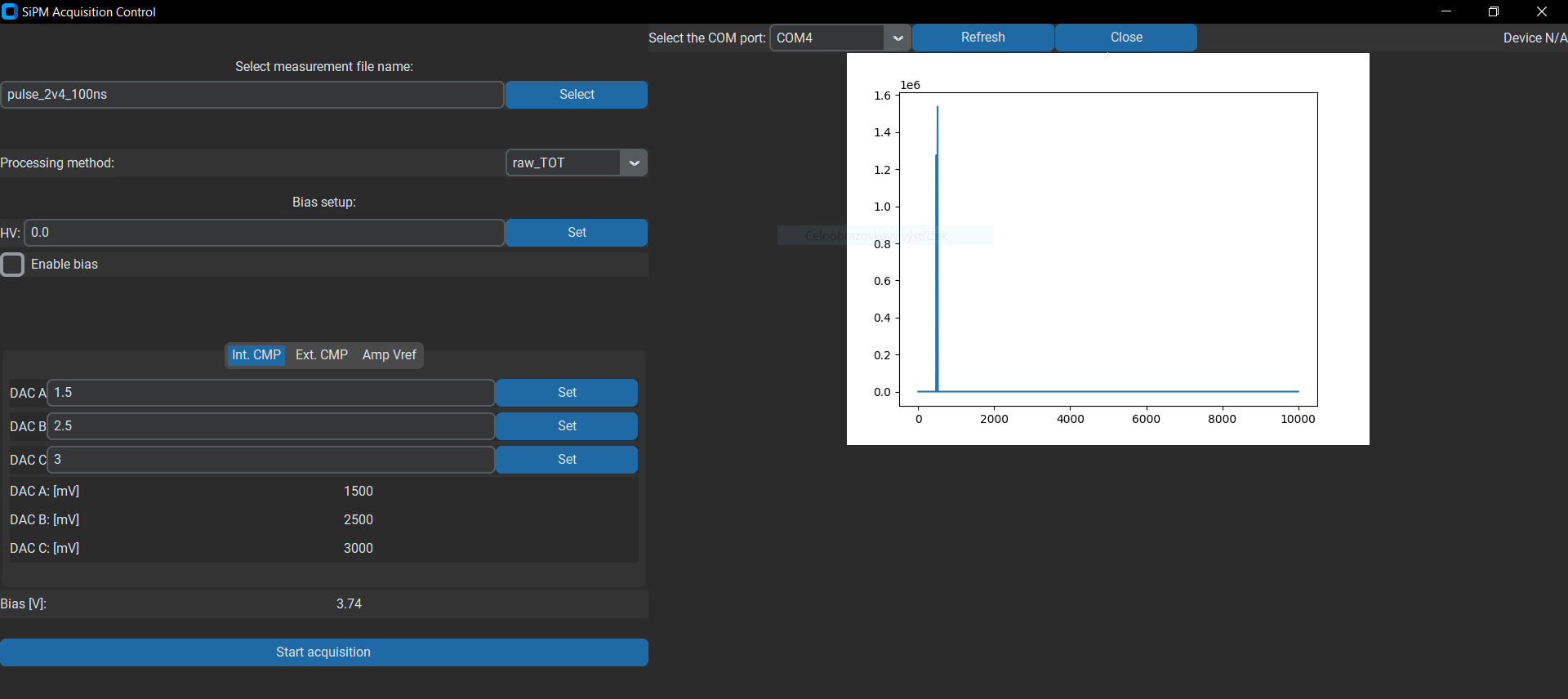}
\qquad
\caption{\label{fig:app} Acquisition software for the MCA - real-time data display view and acquisition setting}
\end{figure}

The interface allows independent configuration of DACs for internal and external comparator thresholds and adjustment of the amplifier offset voltage to account for SiPM breakdown voltage differences. A graphical display provides real-time feedback and detailed control of SiPM settings, as shown in figure ~\ref{fig:app}.

\section{Advanced ToT Measurement}
\label{sec:ToTMeasurement}



The MCA provides three detector inputs with configurable thresholds. Reconfiguration circuitry allows these inputs to combine into a single channel with three independent thresholds, enabling advanced ToT processing for improved energy resolution~\cite{Multi_Voltage_Threshold}.

The system supports two modes: Independent Mode, where data is acquired from three separate channels with individual thresholds for flexible experimentation, and Triple ToT, which uses a single channel to measure signal durations exceeding three thresholds. Triple ToT enhances energy resolution by capturing multiple signal amplitudes, making it ideal for high-temporal-resolution applications.

\section{Spectroscopic Resolution Measurement}
In order to objectively evaluate the properties of the device a series of measurements with a precise signal generator were carried out. Using a signal generator instead of real SiPM detectors coupled with scintillators enables reliable characterization of the device capabilities by neglecting the detector and scintillator properties.

\subsection{Internal Comparator Characterisation}
The key parameter of the comparators in the intended applications is its jitter ~\cite{Jitter_Compact_ASIC,Jitter_ToF_PET}. For use in a more complex measurement chain, where triggering is required signal propagation time is also an important factor~\cite{delay_Nemallapudi_2016,delay_zich,timing_BUZHAN2006353}. Luckily, if the propagation time is known and the jitter is sufficiently low, the propagation delay can often be compensated. Since the device is intended for a balance of low cost and performance, an interesting aspect of the processing chain is especially the internal comparators, embedded in the microcontroller platform. Therefore, the main attention was paid to the jitter measurement of the internal comparators. 

\subsubsection{Comparator Output Pulse Jitter}
The comparator level was set to 1.5 V, representing half of its voltage range, and the pulse duration was measured at 50~\% of its maximum amplitude. The results were then compared with the generator's jitter.

\begin{table}[htbp]
\centering
\caption{\label{tab:jitter}Output pulse properties of the comparators and generator}
\smallskip
\begin{tabular}{ |c|c|c|c|c| } 
\hline
 & Generator & Channel 1 & Channel 2 & Channel 3 \\
\hline
Mean [ns] & 1000.06 & 991.98 & 992.03 & 991.81\\ 
\hline
SD [ps] & 18.3 & 184.3 & 205.8 & 184.2\\ 
\hline
\end{tabular}
\end{table}

The measured jitter of the trigger pulse duration is in table~\ref{tab:jitter} and it is approximately 200 ps. This is a limiting factor for spectroscopy resolution in the case of an external pulse duration measurement carried out ~\cite{delay_Nemallapudi_2016}. In the case of an internal ToT measurement, with the time-step of 184 ps, the jitter falls approximately to the range of one LSB.

\subsubsection{Input to Trigger Output Delay}
As mentioned, the delay from the detector output to the board trigger output, which can be used by other instrumentation, is an important parameter of the device. Significant is the jitter of the delay, which brings uncertainty to the measurement chain. The signal traces from the SiPMs on the PCB are designed to have equal lengths, ensuring consistent signal delay.
In the table ~\ref{tab:Delay} delay analysis results are shown.

\begin{table}[htbp]
\centering
\caption{\label{tab:Delay}Delay of the signal from input to the trigger output connector}
\smallskip
\begin{tabular}{ |c|c|c|c| } 
\hline
 & Channel 1 & Channel 2 & Channel 3 \\
\hline
Mean [ns] & 71.86 & 73.7 & 70.1\\ 
\hline
SD [ps] & 78.81 & 78.81 & 93.13 \\ 
\hline
\end{tabular}
\end{table}

It can be seen that even though the delay is relatively large, the biggest difference between the channels is 3.6 ns, and the greatest measured jitter was 93.13 ps. Since the individual channel delay can be measured and often compensated, and the jitter is relatively low, the triggering system can provide sufficient parameters for other devices in the measurement chain.

\subsubsection{Internal Measurement of Pulse Duration}
\begin{figure}[htbp]
\centering 
\includegraphics[width=0.5\textwidth,trim=0 0 0 0,clip]{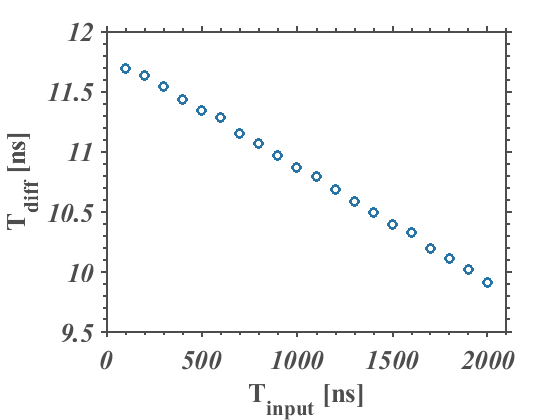}
\qquad
\caption{\label{fig:Tdiff_in_vs_out} Input pulse duration versus absolute error of the acquired duration value}
\end{figure}

A measurement of the input pulses was carried out to measure the resolution of the entire signal processing chain. In this measurement, rectangular pulses of defined duration were injected into the input connector instead of the detector output signal, and the mean values of the pulse and values measured by the microcontroller were compared. 
The results are shown in the figure ~\ref{fig:Tdiff_in_vs_out}.

It can be observed that the absolute error decreases with increasing input pulse duration and has a linear character. The linear character of the error enables its easy correction using calibration.   

\section{Experimental Setup}
\begin{figure}[htbp]
\centering 
\includegraphics[width=0.8\textwidth,trim= 0 0 0 0,clip]{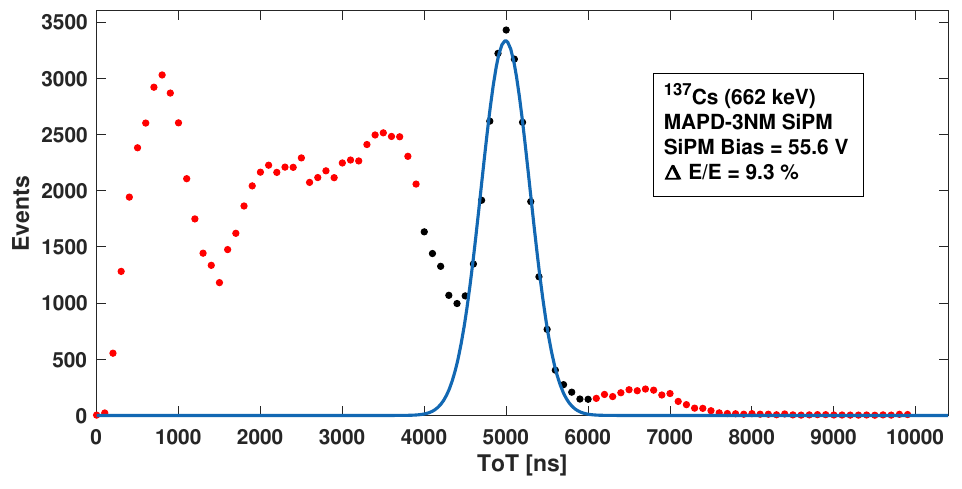}
\qquad
\caption{\label{fig:Histogram_Gauss} Pulse height distribution of $^{137}$Cs (662~keV) from  MAPD-3NM SiPM ~\cite{ahmadov2022investigation} coupled to a LaBr3 scintillator. The SiPM bias was 55.6~V.}


\end{figure}

To validate the performance of the MCA, a series of experiments were conducted using MAPD-3NM silicon photomultiplier ~\cite{ahmadov2022investigation} coupled to a LaBr3 scintillator using a $^{137}$Cs (662~keV) source. The histogram of Pulse height distribution is shown in figure \ref{fig:Histogram_Gauss}. The distribution was measured at room temperature with a bias voltage of 55.6 V. The energy resolution ($\Delta$ E/E), calculated as the Full Width Half Maximum (FWHM) divided by the average pulse height, was obtained by fitting a Gaussian curve to the peak of the distribution. A fit to the distribution, shown in figure \ref{fig:Histogram_Gauss}, yields an energy resolution of 9.3~\%. The pulse height distribution has not been corrected for the effects of dark rate, and crosstalk.

\section{Conclusion}
This paper presents the design and implementation of a compact and cost-effective multi-channel analyzer (MCA) for use with Silicon Photomultipliers (SiPMs) in gamma spectroscopy. By utilizing off-the-shelf components and an STM32G4 microcontroller, the proposed MCA system achieves reliable real-time gamma radiation measurements while maintaining affordability. The device demonstrates good spectroscopic resolution and offers versatile communication options, making it suitable for a wide range of applications, including academic research and educational purposes. The results from experimental tests, using MAPD-3NM SiPM and a LaBr3 scintillator, show that the system provides an energy resolution of 9.3~\%, validating its performance. The MCA's low cost and accessibility expand the opportunities for widespread use in radiation detection. Future work will focus on improving measurement accuracy by calibrating the detectors to real energy values in keV. This enhancement will allow for more precise quantitative analysis in gamma spectroscopy applications.


\acknowledgments

This project has received funding from the European Union’s Horizon 2020 research and innovation programme under the Marie Skłodowska-Curie grant agreement No 101086178, INNMEDSCAN project.



\bibliographystyle{JHEP.bst} 
\bibliography{References.bib}  

\providecommand{\href}[2]{#2}\begingroup\raggedright\begin{thebibliography}{10}

\bibitem{tancioni2021gammadoseratemonitoring}
P.~Tancioni and U.~Gendotti, \emph{Gamma dose rate monitoring using a silicon photomultiplier-based plastic scintillation detector},  2021.

\bibitem{Buonanno2023}
L.~Buonanno, \emph{Gamma-ray spectroscopy and imaging with sipms readout of scintillators: Front-end electronics and position sensitivity algorithms},  in \emph{Special Topics in Information Technology}, C.G.~Riva, ed., (Cham), pp.~41--51, Springer International Publishing (2023), \href{https://doi.org/10.1007/978-3-031-15374-7_4}{DOI}.

\bibitem{david_p__mcelroy__2007}
D.P.~McElroy, V.~Saveliev, A.~Reznik and J.A.~Rowlands, \emph{1. evaluation of silicon photomultipliers: A promising new detector for mr compatible pet}, \href{https://doi.org/10.1016/J.NIMA.2006.10.040}{\emph{Nuclear Instruments \& Methods in Physics Research Section A-accelerators Spectrometers Detectors and Associated Equipment} (2007) }.

\bibitem{ahmadov2022investigation}
F.~Ahmadov, G.~Ahmadov, R.~Akbarov, A.~Aktag, E.~Budak, E.~Doganci et~al., \emph{Investigation of parameters of new mapd-3nm silicon photomultipliers}, {\emph{Journal of Instrumentation} {\bfseries 17} (2022) C01001}.

\bibitem{Holik2020_read-out}
M.~Holik, F.~Ahmadov, G.~Ahmadov, R.~Akbarov, D.~Berikov, Y.~Mora et~al., \emph{Miniaturized read-out interface “spectrig mapd” dedicated for silicon photomultipliers}, \href{https://doi.org/10.1016/j.nima.2020.164440}{\emph{Nuclear Instruments \& Methods in Physics Research, Section A: Accelerators, Spectrometers, Detectors and Associated Equipment} {\bfseries 978} (2020) }.

\bibitem{paolo_trigilio__2018}
P.~Trigilio, P.~Busca, R.~Quaglia, M.~Occhipinti and C.~Fiorini, \emph{A sipm-readout asic for spect applications}, \href{https://doi.org/10.1109/TRPMS.2018.2856201}{\emph{IEEE Transactions on Radiation and Plasma Medical Sciences} {\bfseries 2} (2018) 404}.

\bibitem{SiPM:SADYGOV200670}
Z.~Sadygov, A.~Olshevski, I.~Chirikov, I.~Zheleznykh and A.~Novikov, \emph{Three advanced designs of micro-pixel avalanche photodiodes: Their present status, maximum possibilities and limitations}, \href{https://doi.org/https://doi.org/10.1016/j.nima.2006.05.215}{\emph{Nuclear Instruments \& Methods in Physics Research Section A: Accelerators, Spectrometers, Detectors and Associated Equipment} {\bfseries 567} (2006) 70}.

\bibitem{TI:current_injection}
Texas Instruments, \emph{SLVA861 - How to Dynamically Adjust Power Module Output Voltage}, 12, 2016.

\bibitem{Multi_Voltage_Threshold}
N.~{D'Ascenzo}, L.~{Wang}, X.~{Zhang}, Q.~{Lv}, E.~{Antonecchia}, Y.~{Lin} et~al., \emph{{The JOINBON SiPM for the readout of LySO crystals: a Multi Voltage Threshold approach}}, \href{https://doi.org/10.1088/1748-0221/15/07/C07006}{\emph{Journal of Instrumentation} {\bfseries 15} (2020) C07006}.

\bibitem{Jitter_Compact_ASIC}
S.~Katourani, M.~Besrour, T.~Omrani, K.~Koua, M.~Benhouria, G.~Giustolisi et~al., \emph{Lucas: A low-power ultra-low jitter compact asic for sipm targetting tof-ct},  in \emph{2023 IEEE Nuclear Science Symposium, Medical Imaging Conference and International Symposium on Room-Temperature Semiconductor Detectors (NSS MIC RTSD)}, pp.~1--1, Nov, 2023, \href{https://doi.org/10.1109/NSSMICRTSD49126.2023.10338565}{DOI}.

\bibitem{Jitter_ToF_PET}
M.-A.~Tétrault, A.~Corbeil~Therrien, W.~Lemaire, R.~Fontaine and J.-F.~Pratte, \emph{Tdc array tradeoffs in current and upcoming digital sipm detectors for time-of-flight pet}, \href{https://doi.org/10.1109/TNS.2017.2665878}{\emph{IEEE Transactions on Nuclear Science} {\bfseries 64} (2017) 925}.

\bibitem{delay_Nemallapudi_2016}
M.~Nemallapudi, S.~Gundacker, P.~Lecoq and E.~Auffray, \emph{Single photon time resolution of state of the art sipms}, \href{https://doi.org/10.1088/1748-0221/11/10/P10016}{\emph{Journal of Instrumentation} {\bfseries 11} (2016) P10016}.

\bibitem{delay_zich}
J.~Zich, V.~Georgiev, M.~Holik, V.~Pavlicek and O.~Vavroch, \emph{Multichannel coincidence circuit with settable threshold level for tof afp detector},  in \emph{2019 27th Telecommunications Forum (TELFOR)}, pp.~1--4, Nov, 2019, \href{https://doi.org/10.1109/TELFOR48224.2019.8971310}{DOI}.

\bibitem{timing_BUZHAN2006353}
P.~Buzhan, B.~Dolgoshein, E.~Garutti, M.~Groll, A.~Karakash, V.~Kaplin et~al., \emph{Timing by silicon photomultiplier: A possible application for tof measurements}, \href{https://doi.org/https://doi.org/10.1016/j.nima.2006.05.142}{\emph{Nuclear Instruments \& Methods in Physics Research Section A: Accelerators, Spectrometers, Detectors and Associated Equipment} {\bfseries 567} (2006) 353}.

\end{thebibliography}\endgroup

\end{document}